\documentclass[prl,twocolumn,showpacs,amsmath,amssymb,superscriptaddress,bibnotes,floatfix]{revtex4-1}
\usepackage{graphicx,amsmath,slashbox,bm, braket, txfonts, color}
\usepackage{notes2bib}
\pdfoutput=1

\begin{document}

\title{Theory of electromechanical coupling in dynamical graphene}
\author{Mircea Trif}
\affiliation{Department of Physics and Astronomy, University of California, Los Angeles, California 90095, USA}
\author{Pramey Upadhyaya}
\affiliation{Department of Electrical Engineering, University of California, Los Angeles, California 90095, USA}
\author{Yaroslav Tserkovnyak}
\affiliation{Department of Physics and Astronomy, University of California, Los Angeles, California 90095, USA}

\date{\today}

\begin{abstract}
We study the coupling between  mechanical motion and Dirac electrons in a dynamical sheet of graphene. We show that this coupling can be understood in terms of an effective gauge field acting on the electrons, which has two contributions: {\it quasistatic} and purely {\it dynamic} of the Berry-phase origin. As is well known, the static gauge potential is odd in the $K$ and $K'$ valley index, while we find the dynamic coupling to be even.  In particular, the mechanical fluctuations can thus mediate an indirect coupling between charge and valley degrees of freedom.
\end{abstract}


\maketitle

Graphene has become one of the most remarkable existing material: While the underlying theory is nonrelativistic, its  electronic quasiparticles appear to exhibit relativistic features  as a consequence of the low-energy Dirac spectrum \cite{NovoselovScience04,NovoselovN05, ZhangN05,NovoselovNatMat07,NetoRMP09}. This makes the emergent Dirac fermions Lorentz rather than Galilean invariant \cite{KotovRMP12}. However, this is extremely fragile, as any coupling  beyond the effective model breaks this Lorentz symmetry, as does, for example, Coulombic interaction between electrons or scattering of Dirac electrons by localized impurities. Indeed, a more complete theory must, to a good approximation, be Galilean invariant, which is made explicit if we consider not only the electrons in the frame of reference of a perfect static lattice but the entire graphene membrane as a combined dynamical system.

One prominent feature of the dynamical graphene is lattice dragging \cite{SundaramPRB99}, which could be understood in terms of the Berry phases that electrons accumulate when moving through a time-dependent lattice. The electron dragging  due to lattice translations depends on the difference between the bare electron mass $m$ and the effective electron mass in the crystal $m^*$ \cite{SundaramPRB99}. The larger the difference the stronger the dragging, such that these effects should certainly be relevant in graphene, where the effective electron mass vanishes \cite{WallacePR47}. More generally, such Berry phases can be acquired via both translations and rotations of the local atomic orbitals through an inertial (laboratory) frame of reference: dynamics that may be expected to be significant in free-standing graphene flakes.

The Berry-phase mechanism involving  orbital rotations, while less important in bulk semiconductors, due to the crystalline rigidity, turns out to be of relevance in graphene. This two-dimensional (2D) material is embedded in a three-dimensional (3D) space, which allows for large out-of-plane displacements \cite{VozmedianoPR10} (limited by surface tension and a weak bending rigidity) and thus strong dragging-like effects via rotations of the orbitals. Despite its single-layer structure,  graphene is found to be one of the strongest existing materials \cite{LeeS08,*PootAPL08,*FrankJVSTB07,*BunchScience07,*ChenNatNano09}, with a high degree of elastic control due to its reduced dimensionality. The electronic properties, in turn, are sensitive to the mechanical distortions \cite{PereiraPRL09,GuineaNatMat10,MedvedyevaPRB11,KimPRB11}, offering new pathways for  controlling both the charge and valley  degrees of freedom. In this Letter, we study coupling between the  Dirac electrons  and the time-dependent elastic distortions in  graphene, for a general 3D excitation of the  lattice.


The crystallographic structure of graphene is that of a 2D honeycomb lattice of sp$^2$ hybridized carbon atoms, with two inequivalent lattice sites per unit cell (triangular Bravais lattice with two basis points, $A$ and $B$). The electronic properties of graphene are mainly due to the $p_z$ orbitals of the carbon atoms, which are perpendicular to the graphene sheet and give rise to the so-called $\pi$ band. The other orbitals (sp$^2$ hybridized)  give rise to  the $\sigma$ bands, whose excitations are much higher in energy and  the strong hybridization is responsible  for the mechanical properties of graphene. The approximate tight-binding Hamiltonian describing the 2D graphene accounts for  hopping between only the nearest-neighbor lattice sites  and  reads \cite{WallacePR47}:
\begin{equation}
H_{G}=\sum_s\sum_{p=1,2,3}\left[t_p(\bm{R}_s)c_{\bm{R}_s}^{\dagger}c_{\bm{R}_s+\bm{b}_p}+{\rm H.c.}\right]\,,
\label{Ham}
\end{equation}
where $c^{\dagger}_{\bm{R}}$ ($c_{\bm{R}}$) are the fermionic creation (annihilation) operators at site $\bm{R}$, index $s$ runs over all atom $A$ sublattice sites, $t_p(\bm{R})$ is the hopping amplitude to atom $A$ at position $\bm{R}$ from its neighbor $B$ atoms located at $\bm{R}+\bm{b}_p$, with $\bm{b}_1=(\sqrt{3}/2,1/2)a$, $\bm{b}_2=(-\sqrt{3}/2,1/2)a$, and $\bm{b}_3=(0,-1)a$, $a$ being the equilibrium interatomic distance and $i$ running over all $A$ sites. For  the undistorted graphene, $t_p(\bm{R}_s)\equiv t_0$, i.e., all tunneling matrix elements are equal, which allows one to find the spectrum easily by  writing Hamiltonian (\ref{Ham}) in  the reciprocal space: $c_{\bm{R},\alpha}=\sum_{\bm{k}}c_{\bm{k},\alpha}\exp{(i\bm{k}\cdot\bm{R})}/\sqrt{N}$, where $\alpha$ labels sublattice $A$ or $B$, and $N$ is the total number of unit cells. The reciprocal space of graphene has hexagonal Brillouin zone, with two inequivalent points  $K$ and $K'$ at its vertices. The resultant low-energy spectrum consist of two Dirac cones located at the $K$ and $K'$ points, around which the  spectrum is linear in  momentum $\bm{k}$, $E(\bm{k})=\pm v_F|\bm{k}|$, with $v_F\equiv 3t_0a^2/2$ being the Fermi velocity \cite{WallacePR47}. Transforming back to the real space description,  the effective Hamiltonian around one of those points, say  $K$ point at $\left(4\pi/(3\sqrt{3}a),0\right)$, is  given by (putting $\hbar=1$) $H_{G}^K=-iv_F\bm{\sigma}\cdot\bm{\nabla}$, where $\bm{\sigma}$ is a vector of Pauli matrices acting in the $(A,B)$ sublattice basis. Since the $K'$ point is related to $K$ point by time reversal \cite{NetoRMP09}, their spectra, which define two \textit{valleys,} are essentially identical. This Hamiltonian is responsible for many of  the exotic electronic properties of graphene, such as Klein tunneling \cite{KastneltsonNP06}, the half-integer quantum Hall effect \cite{NovoselovN05, ZhangN05}, {\it zitterbewegung} \cite{Itzykson} etc.

As a starting point, we generalize the above Dirac Hamiltonian to nonuniform (but still real-valued) hoppings $t_p(\bm{R}_i)\neq t_0$. When for any position $\bm{R}_s$, the hopping  $t_p(\bm{R}_s)\equiv t_p$, but $t_p\neq t_{p'}$ for $p\neq p'$, we obtain the following Hamiltonian \cite{VozmedianoPR10}:
\begin{equation}
H_{G}'=v_F\bm{\sigma}\cdot(-i\bm{\nabla}+\bm{A})\,,
\label{StrainedHam}
\end{equation}
with $A_x=(1/3a)(t_1+t_2-2t_3)/t_0$ and $A_y=(1/\sqrt{3}a)(t_2-t_1)/t_0$ being the fictitious gauge potentials emerging as a consequence of the anisotropy in the hopping parameters $t_p$. According to the time-reversal invariance, which is unaffected by a static deformation, the gauge fields have opposite signs at the $K'$ point. Making these gauge fields  space-dependent, the Dirac electrons are subjected to a fictitious magnetic field $\bm{B}_{K,K'}=\pm\bm{\nabla}\times\bm{A}$, with opposite signs in the two valleys.

There  are two main physical mechanisms that lead to modifications in the hopping $t_p(\bm{R}_s)$ in static graphene: due to changes in distance between the carbon atoms and  due to reorientations of the atomic orbitals in a 3D-deformed graphene sheet. In the first case,  the change in the hopping parameter $t_p(\bm{R}_s)$, written in the WKB spirit, reads \cite{VozmedianoPR10}:
\begin{equation}
\delta t_p^{(\beta)}(\bm{R}_s)\simeq t_0\left\{e^{-\beta\left[a_p(\bm{R}_s)/a-1\right] }-1\right\}\approx -\beta t_0\left[a_p(\bm{R}_s)/a-1\right]\,,
\label{beta_strain}
\end{equation}
where $a_p(\bm{R}_s)$ is the modified interatomic distance between the atoms at $\bm{R}_s$ and $\bm{R}_s+\bm{b}_p$. To the leading order, $a_p(\bm{R}_s)/a-1\approx b_p^ib_p^je^s_{ij}/a^2$ (here and henceforth summing over repeated indices $i,j,k,\dots$ that run over the $x,y$ coordinates), where $e_{ij}^s\equiv e_{ij}(\bm{R}_s)$ is the strain tensor at the position $\bm{R}_s$, defined by $e_{ij}=(\partial_iu_j+\partial_ju_i)/2+(\partial_ih)(\partial_jh)$,  with $\bm{u}$ being the in-plane and $h$ the out-of-plane displacements. This expression is independent of the type of orbitals we are dealing with, be it $s$, $p$, $d$, etc. The $\beta$ parameter, however, depends on the orbital, and for graphene $\beta\approx2$ for the $p_z$ orbitals \cite{NetoRMP09}. The resultant gauge field components read $A^{(\beta)}_i=\beta\epsilon_{ij}K_{jkl}e_{kl}/a$, where $K_{xxx}=1$, $K_{xyy}=K_{yxy}=K_{yyx}=-1$ are nonzero elements of the rank-3 trigonal tensor \cite{VozmedianoPR10} and $\epsilon_{ij}$ is the rank-2 Levi-Civita tensor whose nonzero elements are $\epsilon_{xy}=-\epsilon_{yx}=1$. Written explicitly, the vector potential becomes \cite{VozmedianoPR10}:
\begin{equation}
\bm{A}^{(\beta)}(\bm{r})=\frac{\beta}{2a}\left[e_{xx}(\bm{r})-e_{yy}(\bm{r}),-2e_{xy}(\bm{r})\right]\,,
\label{A_stat}
 \end{equation}
which for a nonzero magnetic field thus requires an inhomogeneous strain. Note that we are defining the strain tensor and, subsequently, curvature characterisics of a graphene sheet, along with all gauge fields, with respect to its intrinsic flat coordinate system (i.e., in the graphene's frame of reference).

\begin{figure}[t]
\begin{center}
\includegraphics[width=0.9\linewidth]{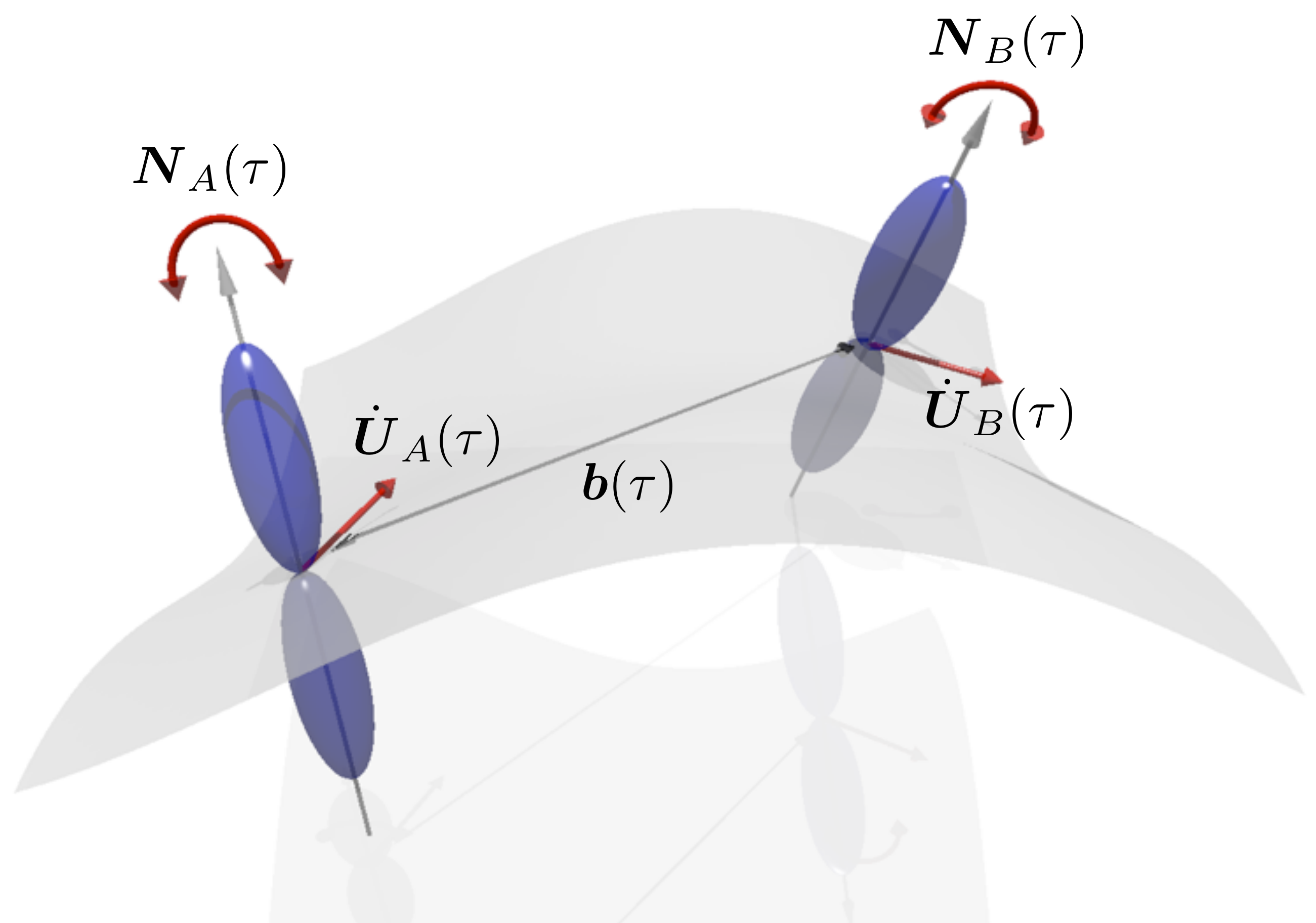}
\caption{(color online). The sketch of a  two-atom ($A$ and $B$) unit-cell basis of a dynamical graphene lattice.  The $p_z$ orbitals (blue dumbells) instantaneously point along the directions $\bm{N}_{A,B}(\tau)$ normal to the graphene substrate (gray), which rotate at angular velocities $\dot{\bm{N}}_{A,B}(\tau)$. Here, $\bm{b}(\tau)$ is the instantaneous distance between the two atoms, and $\bm{\dot{U}}_{A(B)}(\tau)$ are their linear velocities.}
\label{orbitals_sketch}
\end{center}
\end{figure} 

The second mechanism affecting hopping strength involves only the out-of-plane displacement $h(\bm{r})$. As already mentioned,  such distortions cause rotations of the atomic $p_{z}$ and the hybridized sp$^2$ orbitals, which become intermixed, thus modifying the  tunneling Hamiltonian \cite{NetoRMP09}. To quantify the amount of orbital mixture, we introduce the normal to the surface and the unit vector connecting neighboring atoms at the position $\bm{R}_s$, respectively \cite{KimEPL08}:
\begin{align}
\bm{N}_s=\frac{\hat{\bm{z}}-\bm{\nabla} h(\bm{R}_s)}{\sqrt{1+[\bm{\nabla}h(\bm{R}_s)]^2}}\,,\,\,\,
\bm{d}_{s,p}=\frac{\bm{b}_p+\hat{\bm{z}}\delta h_p(\bm{R}_s)}{\sqrt{a^2+[\delta h_p(\bm{R}_s)]^2}}\,,
\end{align}    
where $\delta h_p(\bm{R}_s)\equiv h(\bm{R}_s+\bm{b}_p)-h(\bm{R}_s)$. In the flat graphene, we can write \cite{BardeenPRL61} $t_0=\langle p_{z,A}|H-E_0|p_{z,B}\rangle\approx-2.8\,{\rm eV}$ and $t_0'=\langle p_{y,A}|H-E_0|p_{y,B}\rangle\approx8.4\,{\rm eV}$, where $\bm{R}_B=\bm{R}_A+\bm{b}_3$, $H$ is the full microscopic single-particle Hamiltonian in the lattice, and $E_0$ is the energy corresponding to the $p$ orbitals in an isolated carbon atom. In a curved graphene sheet, this  is generalized to $t_p(\bm{R}_s)=\langle\bm{N}_s\cdot\bm{p}_s|H-E_0|\bm{N}_p\cdot\bm{p}_p\rangle$, with $\bm{p}=(p_{x}, p_{y}, p_{z})$ being the $p$-orbital wave functions at a respective position in the lattice, $\bm{R}_s$ or $\bm{R}_p\equiv \bm{R}_s+\bm{b}_p$. We thus get for the change in tunneling due to curvature:
\begin{align}
&\delta t_p^{(\pi)}(\bm{R}_s)=t_0(\bm{N}_s\cdot\bm{N}_p-1)+(t_0'-t_0)(\bm{N}_s\cdot\bm{d}_{s,p})(\bm{N}_p\cdot\bm{d}_{s,p})\nonumber\\
&\approx-\frac{t_0}{2}\left[(\bm{b}_p\cdot\bm{\nabla})\bm{\nabla} h(\bm{R}_s)\right]^2-\frac{t_0'-t_0}{4a^2}\left[(\bm{b}_p\cdot\bm{\nabla})^2h(\bm{R}_s)\right]^2\,.
\label{t_curv}
\end{align}
In this case too the changes in the tunneling $t_p$ lead to  a fictitious gauge potential $A^{(\pi)}_i=2\sum_p\epsilon_{ij}b_{p}^{j}\delta t^{(\pi)}_p/3t_0a^2$. Written explicitly, the gauge field components $A^{(\pi)}_i(\bm{r})$ read:
\begin{equation}
A^{(\pi)}_{i}(\bm{r})=\frac{3}{8}\epsilon_{ij}\left(K_{jkl}\delta_{mn}+\frac{\mathfrak{t}}{3}K_{jkmln}\right)a\partial_{km}h(\bm{r})\partial_{ln}h(\bm{r})\,,
\label{As}
\end{equation}
where $\mathfrak{t}\equiv(t_0'-t_0)/2t_0$ and the rank-5 tensor $K_{jkrsn}$ has the following nonzero elements: $K_{yyyyy}=-5$, $K_{yxxxx}=\dots=K_{xxxxy}=3$, and $K_{xxyyy}=K_{xyxyy}=\dots=K_{yyyxx}=1$. Expanding this out, say, for the $A^{(\pi)}_x$ component:
\begin{align}
A^{(\pi)}_x=&\frac{3(1+\mathfrak{t})}{8}a\left[(\partial^2_xh)^2-(\partial_y^2h)^2\right]\nonumber\\
&-\frac{\mathfrak{t}}{4}a\left[(\partial_y^2h)^2-(\partial^2_xh)(\partial^2_yh)-2(\partial_{xy}h)^2\right]\,.
\label{Api_x}
\end{align}
We identify  the first term, $\propto(1+\mathfrak{t})$, with the effective gauge field derived  in Ref.~\onlinecite{KimEPL08}, while the second term, $\propto\mathfrak{t}$, is a new contribution that was previously overlooked in the literature (similarly, we obtain 2 terms for $A^{(\pi)}_y$) \cite{note1}.

One could try to extend the above static theory to the dynamical case by simply assuming the strain tensor $\hat{e}(\bm{r})$ and the height $h(\bm{r})$ to be time-dependent in the above expressions. The dynamic distortions then give rise to a fictitious electric field  $\bm{E}(\tau)\equiv-\partial_\tau\bm{A}(\tau)$ ($\tau$ stands for real time), in addition to the aforementioned magnetic fields $\bm{B}=\bm{\nabla}\times\bm{A}$. The effect of such fictitious electric fields \cite{note3}  have been studied recently in connection with both electron energy relaxation in carbon nanotubes \cite{OppenPRB09} and electron charge pumping in graphene \cite{LowNL12}. This quasistatic picture alone would, however,  be incomplete, as it neglects Berry-phase effects engendered by the motion of the atomic basis orbitals themselves in the {\it dynamically} distorted lattice. To capture such effects in graphene, we start by deriving the effective electron tunneling Hamiltonian $H_{\rm eff}$ between two isolated $A/B$ sites in a time-dependent framework. 


The full Hamiltonian of an electron in the potential  $V(\bm{r},\tau)$ created by the two atoms, $A$  and $B$, is  $H_{AB}(\tau)=p^2/2m+V(r,\tau)$, with $m$ being the bare electron mass. Typically, one then solves the Schr{\"o}dinger equation  approximatively by choosing  a trial wave function $|\psi(\tau)\rangle=\left[c_A(\tau)|\psi_A\rangle+c_B(\tau)|\psi_B\rangle\right]\exp{(-iE_0\tau)}$, where $|\psi_{A,B}\rangle\equiv|\bm{N}_{A,B}\cdot\bm{p}_{A,B}\rangle$  and  $c_{A,B}$ respectively stand for the atomic wave functions and their amplitudes at time $\tau$, and derive an effective Hamiltonian $H_{\rm eff}$ in the $A,B$ subspace.  However, these single-atom basis wave functions are not instantaneous eigenstates of the two-atom Hamiltonian $H_{AB}$, which has nonzero matrix elements between the states $|\psi_{A,B}\rangle$  and the other  orbitals outside this effective subspace. In the static situation, one can neglect such couplings, as their effect on $H_{\rm eff}$ appears in second order in perturbation theory with respect to the inverse atomic energy splittings,  while, on the other hand,  the direct coupling in the $|\psi_{A,B}\rangle$ subspace is first order and does not suffer from the energy mismatch.  For the time-dependent case, in contrast, the intra-atomic electronic structure is itself perturbed by dynamics, which requires one to systematically account for the associated excursions outside of the effective subspace $|\psi_{A,B}\rangle$.

To capture such higher-order effects, we perform a more general  time-dependent expansion: $|\psi'(\tau)\rangle=|\psi(\tau)\rangle+\sum_{\alpha=A,B;n\notin\bm{p}_\alpha}c_{n,\alpha}(\tau)|n_{\alpha}\rangle\exp{(-iE_n\tau)}$, with $c_{n,\alpha}(\tau)$ , $|n_\alpha\rangle$, and $E_n$ being respectively the amplitude, wave function, and energy of electron in an excited orbital $n$. Substituting this expression into the Schr{\"o}dinger equation, $i\partial_\tau|\psi'(\tau)\rangle=H_{AB}|\psi'(\tau)\rangle$, leads to the following set of differential equations for $c_{n,\alpha}(\tau)$:
\begin{equation}
i\dot{c}_{n,\alpha}(\tau)=\sum_{n',\alpha'}c_{n',\alpha'}(\tau)\langle n_{\alpha}|H_{AB}-E_n-i\partial_\tau|n'_{\alpha'}\rangle e^{i(E_n-E_{n'})\tau}\,.
\end{equation}
We will solve these equations only approximatively, in perturbation theory, assuming the time dependence of $H_{AB}$ is slow on the time scale  $\tau_d\sim\Delta E$, with $\Delta E$ being the typical orbital level splitting, so that the system stays mostly in a state spanned by the $p$ orbitals. Then, taking  $c_{n,\alpha}(\tau)\approx0$ for excited orbitals, and $|c_{A}(\tau)|^2+|c_{B}(\tau)|^2\approx1$, we obtain in leading order for $n\notin p_\alpha$:
\begin{align}
c_{n,\alpha}(\tau)&\simeq-\sum_{\alpha'}c_{\alpha'}(\tau)\frac{\langle n_{\alpha}|H_{AB}-E_n-i\partial_\tau|n_{\alpha'}\rangle}{E_n-E_{0}}e^{i(E_n-E_0)\tau}+F(\tau)\,,
\label{cna}
\end{align}
where $F(\tau)$ is a function that varies slowly on the time scale $\tau_d$. By substituting this expression in the equations for $c_{A,B}(\tau)$ and neglecting all the fast oscillating terms $\propto\exp{[i(E_0-E_n)\tau]}$,  we end up with the effective Schr{\"o}dinger equation  $i\dot{\bm{c}}(\tau)=H_{\rm eff}\bm{c}(\tau)$, where $H_{\rm eff}\simeq t\,\Sigma_x+\delta\tilde{t}\,\Sigma_y$, with $\delta\tilde{t}=\delta\tilde{t}^{(G)}+\delta\tilde{t}^{(\pi)}$ and
\begin{align}
\label{tG}
\delta\tilde{t}^{(G)}(\tau)=&mt(\tau)\dot{\bm{U}}(\tau)\cdot\bm{b}(\tau)\,,\\
\delta \tilde{t}^{(\pi)}(\tau)=&\frac{1}{2}\left(\left\langle \bm{N}_A\cdot\bm{p}_A|\dot{\bm{N}}_B\cdot\bm{p}_B\right\rangle-\left\langle\dot{\bm{N}}_A\cdot\bm{p}_A|\bm{N}_B\cdot\bm{p}_B\right\rangle\right)\,.
\label{tR}
\end{align}
Here, $\bm{U}(\tau)\equiv\bm{u}(\tau)+\hat{\bm{z}}h(\tau)$ is the sum of in-plane and out-of-plane displacements of the center-of-mass of the two atoms, $\bm{b}(\tau)$ is the distance vector connecting the two atoms,  $t(\tau)\equiv t_0+\delta t(\tau)\approx\langle\psi_{A}|p^2/2m+V(\bm{r},\tau)-E_0|\psi_B\rangle$ is the quasistatic tunneling amplitude discussed above [i.e., $\delta t=\delta t^{(\beta)}+\delta t^{(\pi)}$], $\bm{c}(\tau)\equiv(c_A(\tau), c_B(\tau))$,  and $\bm{\Sigma}$ are Pauli matrices that  act in  this $A,B$ basis. Equation \eqref{tG} can be recognized as a local Galilean boost, for the derivation of which it is crucial to take into account corrections stemming from higher-orbital excursions, Eq.~\eqref{cna} \cite{note2}. Eq.~\eqref{tR} is due to rotations of the local atomic $p$ orbitals (see Fig.~\ref{orbitals_sketch}): It emerges in the time-dependent picture as a Berry-phase correction to the tunneling term defined in Eq. (\ref{t_curv}). Note that the Galilean boost term can be re-exponentiated to give $t\rightarrow t'=t\exp{\left(-im\dot{\bm{U}}\cdot\bm{b}\right)}$, as is expected according to the Peierls substitution in the  moving frame of reference.



We are now equipped  to construct the full graphene Hamiltonian in the dynamical case. The new (dynamic) corrections to the tunneling, $\delta\tilde{t}^{(G,\pi)}$, are purely imaginary, while the quasistatic ones, $\delta t^{(\beta,\pi)}$, are purely real, which  allows us to write $t_{\rm tot}=t_0+\delta t+i\delta \tilde{t}$, with $\delta t=\delta t^{(\beta)}+\delta t^{(\pi)}$ and $\delta \tilde{t}=\delta\tilde{t}^{(G)}+\delta\tilde{t}^{(\pi)}$. The resultant effective Hamiltonian around the $K$ point still has the form of Eq~(\ref{StrainedHam}), but with the total gauge field $\bm{A}_{\rm tot}(\bm{r},\tau)\equiv \bm{A}(\bm{r},\tau)+\bm{\tilde{A}}(\bm{r},\tau)$ being the sum of the quasistatic and dynamic contributions, respectively,  where  $A_{i}=2\sum_p\epsilon_{ij}b_p^j\delta t_p/3t_0a^2$ was evaluated already and $\tilde{A}_{i}=2\sum_{p}b_p^i\delta\tilde{t}_p/3t_0a^2$ consists of:
\begin{align}
\label{AG}
\tilde{A}^{(G)}_i\simeq&-ma\left[\dot{u}_i+a\dot{h}\,\partial_ih+\frac{a^2}{2}K_{ijk}\dot{h}\,\partial_{jk}h\right]\,,\\
\tilde{A}^{(\pi)}_i\simeq&-\frac{1}{t_0}\left[s_{0}\delta_{ij}\delta_{kl}+\frac{s'_{0}-s_{0}}{8}K_{ijkl}\right]\partial_l\dot{h}\,\partial_{jk}h\nonumber\\
&+\frac{a}{2t_0}\left[s_{0}K_{ijk}\delta_{lm}+\frac{s'_{0}-s_{0}}{8}K_{ijlkm}\right]\partial_{km}\dot{h}\,\partial_{jl}h\,,
\label{Api}
\end{align}
where $K_{ijsr}=\delta_{ij}\delta_{kl}+\delta_{ik}\delta_{jl}+{\delta_{il}\delta_{jk}}$,  $s_{0}=\langle p_{z,A}|p_{z,B}\rangle$ and $s_{0}'=\langle p_{y,A}|p_{y,B}\rangle$ (taking their values to be at 0.1 and -0.3, respectively) are orbital overlaps between the nearest-neighbor atoms for $\bm{R}_B=\bm{R}_A+\bm{b}_3$. We find that in the presence of both in-plane and out-of-plane time-dependent distortions, the effective Hamiltonian for graphene is complemented by  the new, purely dynamic, gauge fields $\bm{\tilde{A}}^{(G)}$ and $\bm{\tilde{A}}^{(\pi)}$, which form the main result of our  paper. The  leading-order terms $\propto h(\bm{r},\tau)$ in $\bm{\tilde{A}}^{(G,\pi)}$ appear formally as the dynamic analogue of the static $\bm{A}^{(\beta,\pi)}$ terms, respectively, by substituting one partial spatial derivative $\partial_{i}h\rightarrow\partial_\tau h$, with $i=x,y$, in the static fields. The next-order terms are proportional to the (time) derivative of curvature, $\propto\partial_{ij}\dot{h}$, which do not appear in our static expansion. Repeating the analysis for the $K'$ point (either microscopically or by time reversal of the $K$-point solution) shows that we need to substitute $\bm{A}\rightarrow-\bm{A}$, while $\bm{\tilde{A}}$ has the same sign and magnitude as at the $K$ point. This has profound implications on the electron dynamics, as the gauge fields acting at the $K$ and $K'$ valleys are no longer equal in magnitude, and this asymmetry can lead to time-reversal-breaking observable electromagnetic properties, like charge currents and (quantum) Hall effect. 


 \begin{figure}[t]
\begin{center}
\includegraphics[width=0.99\linewidth]{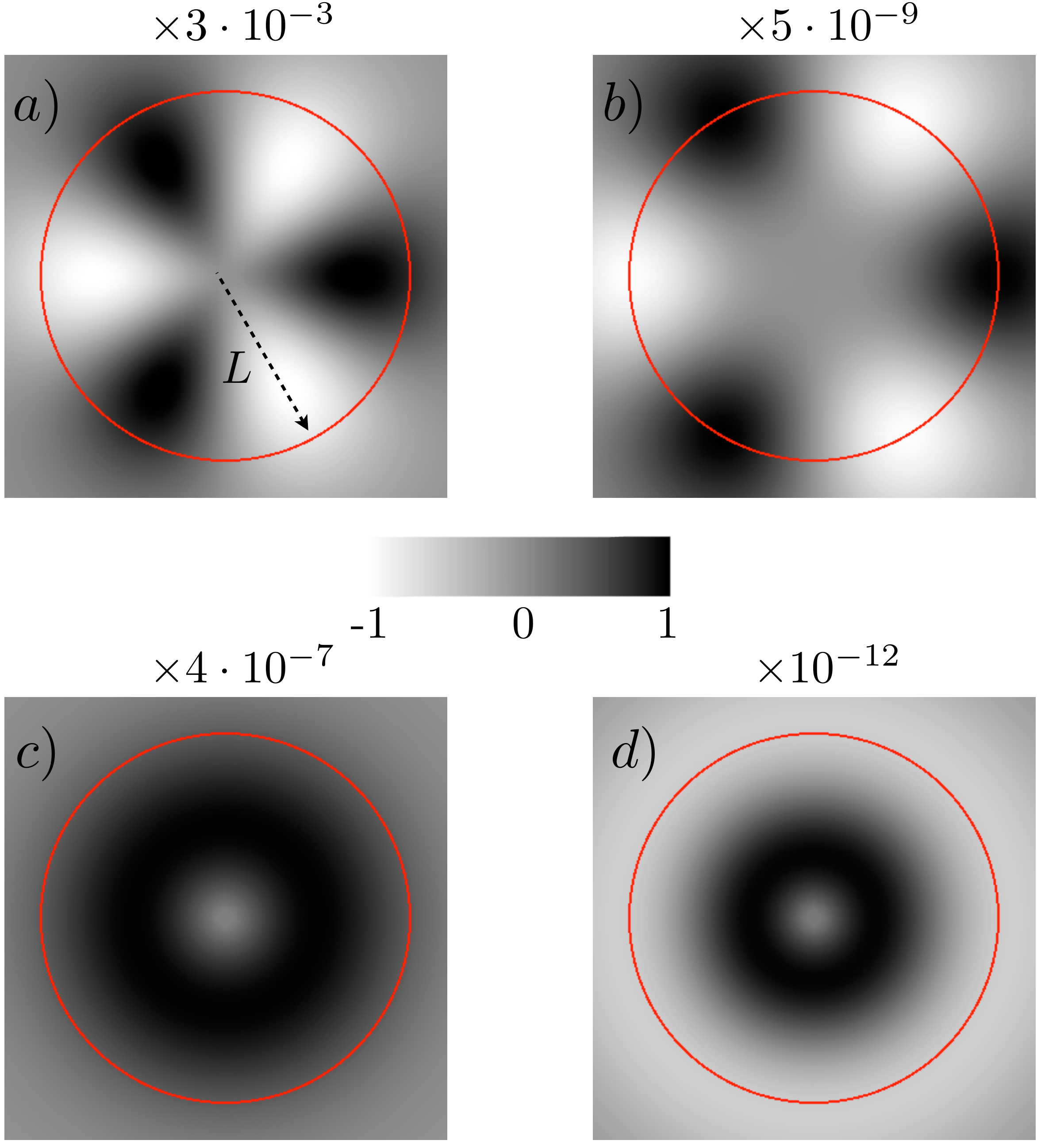}
\caption{The radial projection of the fictitious gauge field, $A_r(\bm{r})\equiv\bm{A}(\bm{r})\cdot\hat{\bm{r}}$, stemming from different  mechanisms discussed in the main text,  for a  Gaussian bump of radius $L=100$ nm, height $h_0=10$ nm and $\omega=10^{9}$~s$^{-1}$. From left to right: a)  {\it static} strain-dependent component  $A_r^{(\beta)}$, b)  {\it static} curvature-dependent component  $A_r^{(\pi)}$, c)  {\it dynamic} Galilean component $\tilde{A}_r^{(G)}$, d) {\it dynamic} curvature-dependent  component $\tilde{A}_r^{(\pi)}$. All the gauge fields are expressed  in units of $a^{-1}$, where $a=1.42$~\AA.}
\label{ArAll}
\end{center}
\end{figure}


As a particular example, we assume a time-dependent gaussian bump, $h(\bm{r},\tau)=h_0\exp{(-r^2/L^2)}\cos{(\omega\,\tau)}$, with $h_0$ and $L$ being the amplitude and the extension of the bump,  respectively. In Fig.~\ref{ArAll}, we plot the resultant gauge field component $A_x$ for the four different mechanisms: strain, Eq.~\eqref{A_stat}, {\it static} curvature, Eq.~\eqref{As},  Galilean boost, Eq.~\eqref{AG}, and {\it dynamic} curvature, Eq.~\eqref{Api}.  We see that for a given Dirac point ($K$ point in Fig.~\ref{ArAll}), the static gauge fields $A_{r}^{(\beta,\pi)}(\bm{r})$ are invariant under $C_3$ rotations,  and under $C_6$ rotations complemented by overall sign change (which corresponds to time reversal), while the dynamic terms $\tilde{A}_{r}^{(G,\pi)}$ are simply invariant under $C_6$ rotations.  Let us now consider the quantitative difference between the four different gauge fields (restoring $\hbar$):
\begin{align}
|\bm{A}^{(\beta)}|\propto \frac{\beta h^2_{\rm 0}}{aL^2}&=\frac{\beta}{a}\left(\frac{h_{\rm 0}}{L}\right)^2\,,\,\,\,
|\bm{A}^{(\pi)}|\propto\frac{ah^2_{\rm 0}}{L^4}=\frac{1}{a}\left(\frac{a}{L}\right)^2\left(\frac{h_{\rm 0}}{L}\right)^2,\nonumber\\
\label{AAA}|\bm{\tilde{A}}^{(G)}|&\propto\frac{m\omega h^{2}_{\rm  0}}{\hbar L}=\frac{1}{a}\frac{\hbar\omega}{E_L}\frac{a}{L}\left(\frac{h_{\rm 0}}{L}\right)^2\,,\\
|\bm{\tilde{A}}^{(\pi)}|&\propto\frac{\hbar\omega h_0^2}{|t_0|L^3}=\frac{1}{a}\frac{\hbar\omega}{|t_0|}\frac{a}{L}\left(\frac{h_{\rm 0}}{L}\right)^2,\nonumber
\end{align}
where $E_L=\hbar^2/mL^2$ is the free-electron energy with a wave number corresponding to the distortion size $L$.
While it may appear that the largest contribution is that arising from strain ($\propto\beta$), it is in fact overestimated here, since graphene is elastically extremely stiff for strain deformations \cite{LeeS08,PootAPL08,FrankJVSTB07,BunchScience07,ChenNatNano09}. Bending, on the other hand, costs much less energy \cite{NetoRMP09}, so that the low-energy height perturbations should favor profiles with minimal strain \cite{DiegoPRB11}. While we treat the planar distortions [setting them to zero in Eq.~\eqref{AAA}] and the height profile $h(\bm{r},\tau)$ as independent parameters, in practice, a self-consistent treatment of elastic properties would be necessary, by balancing bending, shear, and compression energies \cite{KimEPL08}. While addressing this problem in detail is important  in view of quantifying the strength of the emergent gauge fields, such a calculation is beyond our immediate goals and will be presented elsewhere \cite{Pramey2012}. It suffices to say that, due to different spatiotemporal scaling (once the strain and bending are relaxed for a given height profile), we can easily envision geometric and dynamic limits where different terms in Eq.~\eqref{AAA} would provide the dominant contribution.


This work was supported by the NSF under Grant No. DMR-0840965. Fruitful discussions with Antonio H. Castro Neto are gratefully acknowledged.


\begin{thebibliography}{25}
\bibitem{note1}
We note that Eq.~(\ref {Api_x}) reduces to the expression found in Ref.~\onlinecite{KimEPL08} when, for example, $\protect \bm  {\nabla } h\equiv\hat{\bm{x}}\partial _xh$. Geometrically, this means the rotated $p_z$ orbitals are lying in the same $xz$ plane, which simplifies expression for $\delta t^{(\pi)}$ \cite{KimEPL08}. This does not hold for a general noncoplanar case, however, and one needs to use the full expression, Eq.~(\ref {t_curv}), to obtain the correct gauge field $\protect \bm  {A}^{(\pi)}$.
\bibitem{note3}
We note  that there is also a contribution to the electric field due to modulation of the deformation  potential in the distorted lattice, which is given by $\bm{E}_{\rm def}\propto\boldsymbol{\nabla}{\rm Tr}{\hat{e}}$ \cite{AndoPRB02}. However, we disregard these terms here, as they do not give rise to any electromotive forces either over closed circuits or segments of graphene with unstrained boundaries.
\bibitem{note2}
In fact, the result in Eq.~(\ref{tG}) is entirely due  to transitions outside the effective subspace $|\psi_{A,B}\rangle$, via cross terms involving the Hamiltonian $H_{AB}$ and the $\partial_\tau$ operator. The contributions arising from the direct transitions within the $|\psi_{A,B}\rangle$ subspace are exactly cancelled, in a nontrivial way, by a subset of the aforementioned cross terms.
\end{thebibliography}
\end{document}